\documentclass[letterpaper]{biophys_letter}
\usepackage[round, numbers, sort&compress]{natbib}
\usepackage{amsmath,lastpage,bm,textcomp}%mathtime
\usepackage{multicol, float}
\jno{kxl014} %journal number
\gridframe{N}%option for grid around the text "Y" or "N"
\cropmark{N}%option for cropmark around the text "Y" or "N"

\doi{doi: 10.1529/biophysj.000.000000}

\usepackage{amssymb}
\usepackage{graphicx}% Include figure files
\usepackage{verbatim}

\newenvironment{mycols}%
{\begin{multicols}{2}}{\end{multicols}} %double column

\newcommand{\Tscr}{T^*_\mathrm{cr}}
\newcommand{\Rg}{R_{\rm g}}
\newcommand{\Gsum}{\mathcal{G}}

\begin{document}
%
%\titlepage

%$\null$
%\hfill April 18, 2017\\
%$\null$
%\hfill (accepted, {\it Biophysical Journal})\\
%
%\vskip 0.3in
%

\setcounter{page}{1} 
\title{Phase Separation and Single-Chain Compactness of Charged Disordered Proteins are Strongly Correlated}

\author{Yi-Hsuan Lin$^{1,2}$ and Hue Sun Chan$^{1,3}$}
\address{$^1$Department of Biochemistry, University of Toronto,%\\ 
1 King's College Circle, Toronto, Ontario M5S 1A8, Canada; %\\
$^2$Molecular Medicine, Hospital for Sick Children,%\\ 
686 Bay Street, Toronto, Ontario M5G 0A4, Canada; %\\
$^3$Department of Molecular Genetics, University of Toronto,%\\ 
1 King's College Circle, Toronto, Ontario M5S 1A8, Canada}

%\noindent
%$^*$
%Corresponding author information:\\
%Hue Sun C{\footnotesize{HAN}}. $\quad$
%E-mail: chan@arrhenius.med.toronto.edu\\
%Tel: (416)978-2697; Fax: (416)978-8548\\
%Mailing address: Department of Biochemistry, University of Toronto,
%Medical Sciences Building 5207, 1 King's College Circle,
%Toronto, Ontario M5S 1A8, Canada.\\

\maketitle

\pagestyle{headings}

\markboth{Biophysical Journal: Biophysical Letters}{Biophysical Journal: Biophysical Letters}

\begin{abstract}
{Liquid-liquid phase separation of intrinsically 
disordered proteins (IDPs) is a major undergirding factor in the regulated
formation of membraneless organelles in the cell. The phase behavior 
of an IDP is sensitive to its amino acid sequence. Here we apply 
a recent random-phase-approximation polymer theory to investigate
how the tendency for multiple chains of a protein to phase separate, 
as characterized by the critical temperature $T^*_{\rm cr}$, is related 
to the protein's single-chain average radius of gyration $\langle \Rg\rangle$.
For a set of sequences containing different permutations of
an equal number of positively and negatively charged residues, we 
found a striking correlation 
$T^*_{\rm cr}\sim \langle \Rg \rangle^{-\gamma}$
with $\gamma$ as large as $\sim 6.0$,
indicating that electrostatic 
effects have similarly significant impact on promoting
single-chain conformational compactness and phase separation. 
Moreover, $T^*_{\rm cr}\propto -{\rm SCD}$, where SCD is a
recently proposed ``sequence charge decoration'' parameter 
determined solely by sequence information. 
Ramifications of our findings 
for deciphering the sequence dependence of IDP phase separation 
are discussed.}
{Received for publication 1 Jan 0000 and in final form 1 Jan 0000.}
{Address reprint requests and inquiries to H. S. Chan, 
E-mail: chan@arrhenius.med.toronto.edu}
\end{abstract}

\vspace*{2.7pt}
\begin{mycols}
The biological function and disease-causing malfunction of proteins
are underpinned by their structures, dynamics, and myriad intra- and 
inter-molecular interactions. Many critical 
cellular functions are carried out by intrinsically disordered proteins or 
protein regions (collectively abbreviated as IDPs here) with 
sequences that are less hydrophobic than those of globular proteins but are 
enriched in charged, polar, and aromatic 
residues~\cite{Uversky00,tompa12,FormanKay13,vanderLee14,Chen15,Das15}. 
At least 75\% of IDPs are polyampholytes~\cite{Sickmeier07, Das13} 
in that they contain both positively and negatively 
charged residues~\cite{Higgs91, Dobrynin04}. Accordingly, electrostatic 
effects are important in determining individual IDPs' conformational 
dimensions~\cite{Das13, Liu14, Song15} and binding~\cite{Borg07, veronika17}. 
Charge-charge interactions 
are often significant in the recently discovered phenomenon of functional 
IDP liquid-liquid phase separation as 
well~\cite{Wright14, Nott15, Brangwynne15, tanja15, michnick16, pak16, zhang16, feric16}.
IDP phase separation appears to be the physical basis of membraneless 
organelles, performing many vital tasks. Recent
examples include subcompartmentalization within the nucleolus~\cite{feric16}
and synaptic plasticity~\cite{zhang16}. Malfunction of phase separation 
processes can lead to disease-causing amyloidogenesis~\cite{tanja15}
and neurological disorders~\cite{zhang16}. Speculatively, membraneless
liquid-liquid phase separation of biomolecules might even have played 
a role in the origins of life~\cite{keating12acc}.

Electrostatic effects encoded by a sequence of charges 
depend not only on the total positive and negative charges or net 
charge~\cite{Mao10, Muller10} but also the charge 
pattern~\cite{Das13}. For IDPs, this
was demonstrated by Das and Pappu who conducted
explicit-chain, implicit-solvent conformational sampling of thirty 
different sequences each composing of 25 lysine (K) and 25 glutamic acid (E) 
residues (termed KE sequences hereafter). They found that the average 
radius of gyration, $\langle\Rg\rangle$, is strongly sequence-dependent, 
and is correlated with a charge pattern parameter $\kappa$ that
quantifies local deviations from global charge asymmetry~\cite{Das13}.
A subsequent analytical treatment of the KE sequences by Sawle and
Ghosh rationalized the trend through another charge pattern
parameter ``sequence charge decoration'' (SCD) that also correlates
well with $\langle\Rg\rangle$~\cite{Sawle15}. For IDP
phase separation, a recent sequence-dependent random-phase-approximation (RPA) 
approach we put forth~\cite{Lin16a, Lin17} accounted for
the experimental difference in phase-separation tendency
between the wildtype and a charge-scrambled mutant of the 236-residue
N-terminal fragment of DEAD-box RNA helicase Ddx4~\cite{Nott15}.

These advances suggest that a deeper understanding of the 
fundamental relationship between single- and multiple-chain IDP properties 
is in order. It would be helpful, for instance, if experiments
on single-chain properties can infer the conditions under which
a protein sequence would undergo multiple-chain phase separation. 
We embark on this endeavor by first focusing on electrostatics, 
while leaving aromatic and other 
$\pi$-interactions---which can figure prominently in IDP 
behavior~\cite{Nott15, Lin16a, Song13}---to future effort. 
To reach this initial goal, we apply RPA to the
thirty KE sequences of length $N\!=\!50$ to ascertain 
their phase-separation properties under salt-free conditions. Adopting 
our previous notation and making the same simplifying assumption that 
amino acid residues and water molecules are of equal size in the 
theory~\cite{Lin16a, Lin17}, the free energy $F_{\rm RPA}$ of the
multiple-chain system of a given polyampholytic sequence with
charge pattern $\{ \sigma_i \} = \{\sigma_1, \sigma_2 ... \sigma_N\}$, 
%where $\sigma_i = \pm 1$ is the sign of electronic charge of the $i$th residue, is 
%given by [see Eqs.~(13) and (40) of Ref.~\cite{Lin17}]:
\doiline
\end{mycols}

\begin{figure}
\centering{\includegraphics[width=\textwidth]{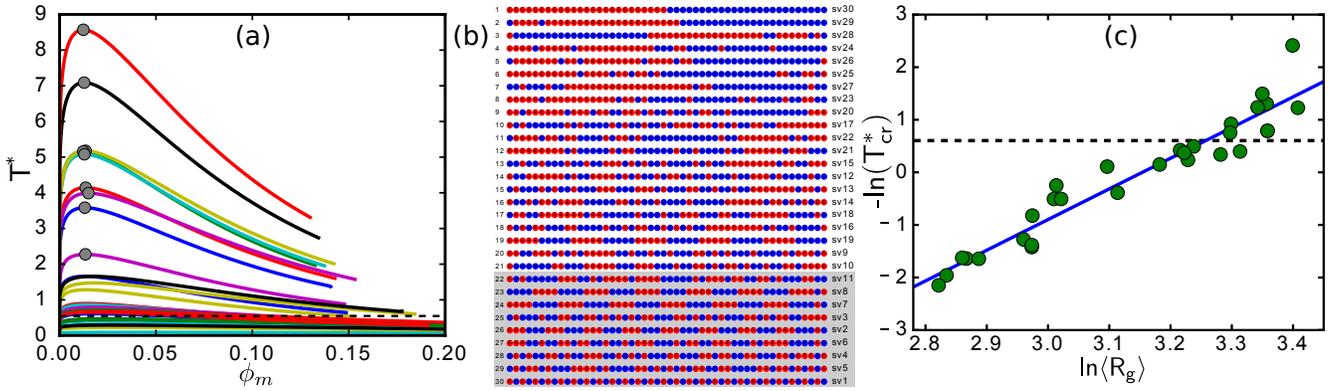}}
\caption{(a) Coexistence curves computed by RPA
for KE sequences 1--30 in (b), listed in descending order of $\Tscr$ 
(except sequences 23 and 24 which have the same $\Tscr$),
with K and E residues in red and blue, respectively;
those with $\Tscr < 0.55$ (corresponding to $T< 300$ K when 
$\epsilon_r=80$) are shown on a grey background in (b).
The ``sv'' sequence labels are those in Ref.~\cite{Das13}.
Critical points ($T^*=\Tscr$) for several high-$\Tscr$ sequences
are marked by circles in (a). 
(c) Logarithmic correlation between RPA-predicted $\Tscr$ and
$\langle\Rg\rangle$ simulated in Ref.~\cite{Das13}
(green circles). The fitted line (blue) is $-\ln\Tscr
= - 18.4 + 5.83\ln\langle \Rg \rangle$ with squared Pearson coefficient 
$r^2 = 0.92$. The dashed horizontal line 
represents $\Tscr = 0.55$.}
	\label{fig:pappu30}
\end{figure}

\begin{mycols}
\noindent
%multiple-chain system of a given polyampholytic sequence with
%charge pattern $\{ \sigma_i \} = \{\sigma_1, \sigma_2 ... \sigma_N\}$, 
where $\sigma_i = \pm 1$ is the sign of electronic charge of the $i$th residue, is 
given by [see Eqs.~(13) and (40) of Ref.~\cite{Lin17}]:
\begin{eqnarray}
\frac{F_{\rm RPA}a^3}{V k_{\rm B} T} & = & \frac{\phi_m}{N}\ln\phi_m 
+ (1-\phi_m)\ln(1-\phi_m) \nonumber \\
&& + \int_0^{\infty} \frac{dk k^2}{4\pi^2}\left\{ \ln\left[ 1+ \Gsum(k)\right] - \Gsum(k) \right\},
\label{eq1}
\end{eqnarray}
where $a=3.8$\AA~is the C$\alpha$-C$\alpha$ distance, 
$V$ is system volume, $k_{\rm B}$ is Boltzmann constant, $T$ is absolute
temperature, $\phi_m = \rho_m a^3$ is the volume ratio of amino residues 
wherein $\rho_m/N$ is protein density, and
\begin{equation}
\Gsum(k) = \frac{4\pi \phi_m}{k^2(1+k^2)T^*N} 
 \! \sum_{i,j=1}^N \! \sigma_i\sigma_j 
\exp\left(-\frac{k^2}{6}|i\!-\!j| \right). 
	\label{eq:G_RPA}
\end{equation}
Here $T^* \equiv a/l_B$ is reduced temperature; the Bjerrum length
$l_B = e^2/(4\pi\epsilon_0 \epsilon_r k_{\rm B} T)$ 
where $e$ is elementary charge, 
$\epsilon_0$ is vacuum permittivity, and $\epsilon_r$ 
is relative permittivity~\cite{Lin16a, Lin17};  $\epsilon_r\approx 80$ 
for water but can be significantly lower for water-IDP solutions~\cite{Lin17}.
Here $\epsilon_r$ is treated largely as an unspecified constant because 
our main concern is the relative $\Tscr$s of different sequences.

We determined the phase diagrams of the 30 KE sequences from the free 
energy expression Eq.~(\ref{eq1}) using standard procedures 
\cite{Lin17}.  For each sequence, the highest temperature on the 
coexistence curve is the critical temperature $\Tscr$, which is the 
highest $T^*$
at which phase separation can occur (Fig.~1(a)). The
critical temperatures of the KE sequences are highly diverse,
ranging from $\Tscr=0.089$ (sv1) to $8.570$ (sv30). The variation
of critical volume fraction $\phi_{\rm cr}\equiv\phi_m(\Tscr)$ 
from $0.0123$ (sv30, sv24) to $0.0398$ (sv1) is narrower. 
The KE sequences were originally
labeled as sv1, sv2, $\dots$, sv30 in ascending values for Das and Pappu's
charge pattern parameter $\kappa$, from the
strictly alternating sequence sv1 with $\kappa=0.0009$ (minimum segregation
of opposite charges) to the diblock sequence sv30 with $\kappa=1.0$
(maximum charge segregation) \cite{Das13}.  Our RPA-predicted
$\Tscr$s follow largely, though not exactly, the same order: sv1 and sv30
have the lowest and highest $\Tscr$s, respectively; but, e.g., sv24 rather than
sv27 has the fourth largest $\Tscr$ and sv5, not sv2, has the
second lowest $\Tscr$. If $\epsilon_r = 80$ is assumed, RPA predicts that
21 KE sequences can, but 9 KE sequences cannot phase separate at 
$T \ge 300$ K (Fig.~1(b,c)).

Because $\langle\Rg\rangle$ correlates positively with $\kappa$ \cite{Das13},
the present $\Tscr$ trend suggests that multiple-chain $\Tscr$ should 
correlate with single-chain $\langle\Rg\rangle$.
Indeed, a striking correlation (Fig.~1(b)) satisfying
the approximate power-law 
\begin{equation}
\Tscr \approx 9.8\times10^7 \langle\Rg\rangle^{-5.83} \; ,
	\label{eq:TcRg_power}
\end{equation}
with $\Rg$ in units of \AA, is observed for the KE sequences. 
The variation of $\Tscr$ with $\langle\Rg\rangle$ is very sharp:
$\Tscr$ increases $\sim 100$ times while
$\langle\Rg\rangle$ decreases by $\lesssim 50\%$.
Qualitatively, the
positive ($\Tscr$)--$\langle\Rg\rangle$ correlation may
be understood by considering two extreme cases:
The diblock and the strictly alternating sequences
(Fig.~2). For the diblock, 
attractive interactions are absent---cannot be satisfied---within
most stretches of several (e.g. $< 6$) residues. 
However, once a pair of opposite charges are in spatial proximity, chain 
connectivity brings two oppositely charged blocks together, leading
to a strong Coulomb attraction, thus a small $\langle\Rg\rangle$ 
and a higher tendency to phase separate 
(higher $\Tscr$). In contrast, for the strictly 
alternating sequence,
attractive Coulumb interactions that are already weakened relative
to that of the diblock sequence require more conformational restriction,
resulting in more open, large-$\langle\Rg\rangle$ single-chain conformations
and less tendency to phase separate (lower $\Tscr$).

It is instructive to compare the predictive power
of $\kappa$ and another charge pattern parameter 
SCD $\equiv \sum_{i<j}^N\sigma_i\sigma_j\sqrt{j-i}/N$ that 
has emerged from the 
analysis of Sawle and Ghosh \cite{Sawle15}. 
The two parameters are well correlated ($r^2=0.95$, see Fig.~7 of Ref.~\cite{Sawle15}), yet 
the variation of both $\Tscr$ and $\langle\Rg\rangle$ of the KE sequences
is significantly smoother with respect to SCD than $\kappa$ 
(Fig.~3). For example, 
\end{mycols}

\twocolumn

\begin{figure}
\centering{\includegraphics[width=\columnwidth]{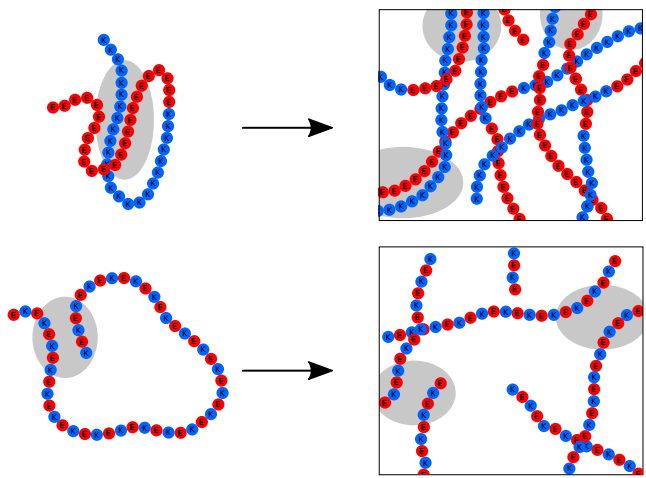}}
\caption{Schematics: similar electrostatic effects
are at play in single-chain compactness (left)
and multiple-chain phase separation (right).
Top: Long stretches of like charges entails
strong intra- and interchain attractions (grey areas). Favorable
intrachain interactions are among residues nonlocal, i.e., more than 
a few residues apart, along the chain sequence. 
Most local interactions are repulsive because of the charge blocks. 
Bottom: Attraction within and among polyampholytes that lack long
charge blocks are weaker. Overall 
attractive interactions now require conformationally restrictive 
charge pairings and are weaker because
of repulsion from neighboring like charges.}
	\label{fig:schematic}
\end{figure}

\noindent
despite the large variation
in $\kappa$ for sv24, sv26, and sv28 
($0.45$, $0.61$, and $0.77$, respectively), 
their $\langle\Rg\rangle=17.6$, $17.5$, 
and $17.9$\AA~\cite{Das13}, and their 
$\Tscr=5.16$, $5.08$, and $5.18$ are
almost identical (Fig.~3(b)). This similarity,
however, is well 
reflected by their similar
SCD $=-17.0$, $-16.2$, and $-16.0$.  Indeed, a near-linear
relationship ($r^2=0.997$), 
\begin{equation}
\Tscr \approx -0.314 ({\rm SCD}) \; ,
\label{eq4}
\end{equation}
is observed (Fig.~3(b)). 
A likely origin of SCD's better performance is that it accounts
for potential interactions between charges far apart along the 
sequence whereas
$\kappa$ relies on averaging over 5 or 6 consecutive charges. Accordingly,
SCD is less sensitive than $\kappa$ to isolated charge reversals.
The rather smooth SCD--$\langle\Rg\rangle$ dependence
is remarkable because the simulated $\langle\Rg\rangle$ \cite{Das13} 
bears no formal relationship with the variational theory from which 
SCD emerges \cite{Sawle15}.
Future effort should be directed toward further assessment of these and 
other possible charge pattern parameters~\cite{cider17} 
as predictors for IDP conformational properties.

\begin{figure}
\centering{\includegraphics[width=\columnwidth]{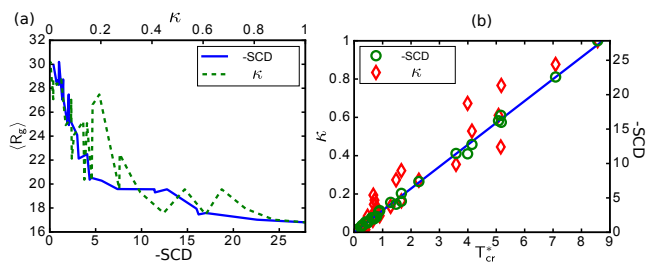}}
\caption{Charge-pattern parameters.
(a) Single-chain $\langle \Rg\rangle$ in \cite{Das13} versus
the $\kappa$ parameter of Das and Pappu ~\cite{Das13} (top horizontal scale)
and the SCD parameter of Sawle and Ghosh~\cite{Sawle15} (bottom scale 
for $-$SCD).
(b) Variation of RPA-predicted $\Tscr$ with $\kappa$ (left vertical scale) and
$-$SCD (right vertical scale).}
	\label{fig:kappa_SCD}	
\end{figure}

In summary, we have quantified a close relationship 
between single-chain conformational compactness of polyampholytes
and their phase separation tendency. The above RPA results 
were derived with a short-range cutoff for 
Coulumb interactions to account for residue 
sizes~\cite{Lin17, Ermoshkin03}. If we had
adopted an unphysical interaction scheme without such a cutoff,
similar trends would still hold although the scaling relations
Eqs.~(\ref{eq:TcRg_power}) and (\ref{eq4}) would be modified, respectively, 
to $\Tscr \sim (\Rg)^{-3.57}$ and $\Tscr \approx -0.490 ({\rm SCD})$.
Thus, in any event, basic physics dictates
a rather sharp positive correlation between $\Tscr$ and $\langle\Rg\rangle$.
This connection should be further explored by both theory and
simulation~\cite{cider17, ruff15} to help decipher the 
sequence determinants of IDP phase separation.

%$\null$

\section*
{AUTHOR CONTRIBUTIONS}

Y.-H.L. and H.S.C. designed research, performed research, analyzed data. 
and wrote the article.

\section*
{ACKNOWLEDGMENTS}

We thank Julie Forman-Kay and Robert Vernon for helpful discussions.
This work was supported by Canadian Cancer Society Research Institute 
grant no. 703477, Canadian Institutes of Health Research grant MOP-84281, 
and computational resources provided by SciNet of Compute Canada.

%\vfill\eject

%\centerline{\large\bf Figure Captions}

%$\null$

%\noindent
%{\bf Figure 1.} 
%\\

%\noindent
%{\bf Figure 2.}
%\\

%\noindent
%{\bf Figure 3.}
%\\

%\vfill\eject

%\section*{References}
%\vspace{-0.2in}

\end{document}